\documentclass[preprint,showpacs,preprintnumbers,amsmath,amssymb,nofootinbib]{revtex4}

\usepackage{epsfig}
\usepackage{graphicx}
\usepackage{dcolumn}
\usepackage{bm}

\def\DESepsf(#1 width #2){\epsfxsize=#2 \epsfbox{#1}}

\begin{document}


\title{Rescattering effects in $\overline B_{u,d,s}\to D P,\overline D P$ decays}
%

\author{Chun-Khiang Chua}
\affiliation{ Physics Department, Chung Yuan Christian University,
Chung-Li, Taiwan 32023, Republic of China
}%
\author{Wei-Shu Hou}%
\affiliation{%
Physics Department, National Taiwan University, Taipei, Taiwan
10764, Republic of China
}%

\date{December 12, 2007}

\begin{abstract}
We study quasi-elastic rescattering effects in $\overline
B_{u,d,s}\to DP$, $\overline DP$ decays, where $P$ is a light
pseudoscalar. The updated measurements of $\overline B_{u,d}\to
DP$ decays are used to extract the effective Wilson coefficients
$a^{\rm eff}_1\simeq 0.90$, $a^{\rm eff}_2\simeq 0.23$, three
strong phases $\delta\simeq 53^\circ$, $\theta\simeq 18^\circ$,
$\sigma\simeq -88^\circ$, and the mixing angle $\tau\simeq
9^\circ$. This information is used to predict rates of nineteen
$\overline B_{s}\to DP$ and $\overline B_{u,d,s}\to \overline DP$
decay modes, including modes of interests in the $\gamma/\phi_3$
program. Many decay rates are found to be enhanced. In particular,
the $\overline B_s\to D^0 K^0$ rate is predicted to be $8\times
10^{-4}$, which could be measured soon. The rescattering effects
on the corresponding $\overline B_{u,d,s}\to\overline DP, DP$
amplitude ratios $r_B$, $r_{B_s}$, and the relative strong phases
$\delta_B$, $\delta_{B_s}$ are studied. Although the decay rates
are enhanced in most cases, $r_{B,B_s}$ values are similar to
factorization expectation.

\end{abstract}

\maketitle

\section{Introduction}

Color-suppressed $b\to c$ decays $\overline B^0\to
D^{(*)0}\pi^0$~\cite{BelleBDpi,CLEOBDpi},
$D^0\eta,D^0\omega$~\cite{BelleBDpi}, $D^0\eta'$~\cite{BaBarBDpi},
$D^+_s K^-$ and $D^0 \overline K{}^0$~\cite{DsK,D0K0} started to
emerge in 2001 (for updated measurements, see \cite{HFAGc,PDG}),
with branching ratios that are significantly larger than earlier
theoretical expectations based on naive factorization. When
combined with color-allowed $\overline B\to D^{(*)}\pi$ modes in
an SU(2) framework, the enhancement in the $D^{(*)0}\pi^0$ rate
indicates the presence of non-vanishing strong phases, which has
attracted much attention~\cite{Xing:2001nj,Cheng:2001sc,
Neubert:2001sj,Chua:2001br,Chiang:2002tv,SCET,pQCD,Gronau:2002mu,
Fleischer:2003,Wolfenstein:2003pc,Cheng:2004ru}. 
We proposed~\cite{Chua:2001br} a quasi-elastic final state
rescattering (FSI) picture, where the enhancement of
color-suppressed $D^0 h^0$ modes is due to rescattering from the
color-allowed $D^+\pi^-$ final state. This approach was also
applied to study final state interaction in charmless $B$
decays~\cite{Chua:2002wk}.

The quasi-elastic approach was recently extended to $\overline
B\to D\overline K$, $\overline D\,\overline K$
decays~\cite{Chua:2005dt}. The color-allowed $B^-\to D^0K^-$ and
color-suppressed $B^-\to \overline D {}^0 K^-$ decays are of
interest for the determination of the unitary phase angle
$\phi_3({\rm or}\ \gamma)\equiv \arg V_{ub}^*$, where $V$ is the
Cabibbo-Kobayashi-Maskawa (CKM) quark mixing matrix. The
Gronau-London-Wyler (GLW)~\cite{GLW}, Atwood-Dunietz-Soni
(ADS)~\cite{ADS} and ``$DK$ Dalitz
plot"~\cite{Dalitz,DalitzBelle0304} methods probe, in varying
ways, the interference of the two types of amplitudes in a common
final state. The enhancement of color-suppressed $DP$ modes (where
$P$ stands for a light pseudoscalar) could imply a larger
$\overline D K$ rate~\cite{Gronau:2002mu}. Since the strong
interaction respects SU(3) and charge conjugation symmetries, the
FSI in $DP$ and $\overline D P$ modes should be related. It is
thus of interest to study $DP$ and $\overline D\,P$ modes
together.

Besides making an update with recently available data, we note
that data for $\overline B_s$ is starting to emerge from the
Tevetron~\cite{PDG} and from $B$ factories~\cite{BelleBsexcl}, and
we anticipate more to come in the near future, from LHCb and other
LHC experiments. Some $B_s$ modes will be useful in the extraction
of $\gamma/\phi_3$~\cite{GL,ADK,Chua:2005fj}. It is thus timely to
study $\overline B_s$ decays.  In this work, we extend the scope
of the quasi-elastic rescattering approach to $\overline B_s\to
DP,\,\overline D P$ decays, as well as update our previous results
using the latest $\overline B_{u,d}\to DP$ data~\cite{HFAGc,PDG}.

In Sec. II we briefly summarize and extend the quasi-elastic
rescattering formula for $\overline B_{u,d,s}\to DP,\,\overline D
P$ decays. Numerical results are reported in Sec.~III. The
effective Wilson coefficients and rescattering parameters are
obtained by using current $\overline B\to DP$ data. By SU(3)
symmetry and charge conjugation invariance of the strong
interactions, we make predictions on $\overline B_s\to D P$ and
$\overline B_{u,d,s}\to \overline D P$ rates. The conclusion is
then offered in Sec. IV.
An Appendix specifies the source amplitudes used to fit data with
rescattering formalism.

\section{Final State Rescattering Framework}


We only briefly summarize, as well as extend, the decay amplitudes
obtained in the quasi-elastic approach for $\overline B\to DP,
\overline D P$ decays, and refer the reader to
Refs.~\cite{Chua:2001br} and~\cite{Chua:2005dt} for more detail.

The quasi-elastic strong rescattering amplitudes can be put in
four different classes, as given below.
For $\overline B$ decaying to $DP$ with $C = +1$, $S = 0,\;-1$
final states, we have
\begin{eqnarray}
A_{B^-\to D^0\pi^-(D^0 K^-)}&=&(1+i r'_0+ir'_e)\, A^0_{B^-\to
D^0\pi^-(D^0 K^-)},
\nonumber\\
\left(
\begin{array}{l}
A_{\overline B {}^0\to D^+ K^-}\\
A_{\overline B {}^0\to D^0 \overline K {}^0}
\end{array}
\right)
 &=& {\cal S}^{1/2}_1
 \left(
\begin{array}{l}
A^0_{\overline B {}^0\to D^+ K^-}\\
A^0_{\overline B {}^0\to D^0 \overline K{}^0}
\end{array}
\right),
 \nonumber\\
 \left(
\begin{array}{l}
A_{\overline B {}^0\to D^+\pi^-}\\
A_{\overline B {}^0\to D^0\pi^0}\\
A_{\overline B {}^0\to D^+_s K^-}\\
A_{\overline B {}^0\to D^0\eta_8}\\
A_{\overline B {}^0\to D^0\eta_1}
\end{array}
\right) &=&{\cal S}^{1/2}_2\, \left(
\begin{array}{l}
A^0_{\overline B {}^0\to D^+\pi^-}\\
A^0_{\overline B {}^0\to D^0\pi^0}\\
A^0_{\overline B {}^0\to D^+_s K^-}\\
A^0_{\overline B {}^0\to D^0\eta_8}\\
A^0_{\overline B {}^0\to D^0\eta_1}
\end{array}
\right).
 \label{eq:FSIBDP}
\end{eqnarray}
Extending to $\overline B_s$ to $DP$ decays with $C = +1$, $S =
0,\;+1$ final states, one has
\begin{eqnarray}
\left(
\begin{array}{l}
A_{\overline B {}^0_s\to D^+_s \pi^-}\\
A_{\overline B {}^0_s\to D^0 K^0}
\end{array}
\right)
 &=& {\cal S}^{1/2}_1
 \left(
\begin{array}{l}
A^0_{\overline B {}^0_s\to D^+_s \pi^-}\\
A^0_{\overline B {}^0_s\to D^0 K^0}
\end{array}
\right),
 \nonumber\\
 \left(
\begin{array}{l}
A_{\overline B {}^0_s\to D^+\pi^-}\\
A_{\overline B {}^0_s\to D^0\pi^0}\\
A_{\overline B {}^0_s\to D^+_s K^-}\\
A_{\overline B {}^0_s\to D^0\eta_8}\\
A_{\overline B {}^0_s\to D^0\eta_1}
\end{array}
\right) &=&{\cal S}^{1/2}_2\, \left(
\begin{array}{l}
A^0_{\overline B {}^0_s\to D^+\pi^-}\\
A^0_{\overline B {}^0_s\to D^0\pi^0}\\
A^0_{\overline B {}^0_s\to D^+_s K^-}\\
A^0_{\overline B {}^0_s\to D^0\eta_8}\\
A^0_{\overline B {}^0_s\to D^0\eta_1}
\end{array}
\right).
 \label{eq:FSIBsDP}
\end{eqnarray}
For $\overline B {}_{u,d}\to \overline D P$ decays with $C = -1$,
$S = \pm1$ final states,
\begin{eqnarray}
 \left(
\begin{array}{l}
A_{\overline B {}^0\to D^- K^+}\\
A_{\overline B {}^0\to \overline D^0 \overline K {}^0}
\end{array}
\right)
 &=& {\cal S}^{1/2}_1
 \left(
\begin{array}{l}
A^0_{\overline B {}^0\to D^- K^+}\\
A^0_{\overline B {}^0\to D^0 \overline K{}^0}
\end{array}
\right),
 \nonumber\\
\left(
\begin{array}{l}
A_{B^-\to\overline D {}^0 K^-}\\
A_{B^-\to D^- \overline K{}^0}\\
A_{B^-\to D^-_s \pi^0}\\
A_{B^-\to D^-_s\eta_8}\\
A_{B^-\to D^-_s\eta_1}
\end{array}
\right)
 &=& {\cal S}^{1/2}_3\, \left(
\begin{array}{l}
A^0_{B^-\to \overline D {}^0 K^-}\\
A^0_{B^-\to D^- \overline K{}^0}\\
A^0_{B^-\to D^-_s \pi^0}\\
A^0_{B^-\to D^-_s\eta_8}\\
A^0_{B^-\to D^-_s\eta_1}
\end{array}
\right).
 \label{eq:FSIBDbarP}
 \end{eqnarray}
And for $\overline B {}^0_s\to \overline DP$ decays with $C = -1$,
$S = 0,\;+1$ final states,
\begin{eqnarray}
 \left(
\begin{array}{l}
A_{\overline B {}^0_s\to D^-\pi^+}\\
A_{\overline B {}^0_s\to \overline D^0\pi^0}\\
A_{\overline B {}^0_s\to \overline D^-_s K^+}\\
A_{\overline B {}^0_s\to \overline D^0\eta_8}\\
A_{\overline B {}^0_s\to \overline D^0\eta_1}
\end{array}
\right)
 &=&{\cal S}^{1/2}_2\, \left(
\begin{array}{l}
A^0_{\overline B {}^0_s\to D^-\pi^+}\\
A^0_{\overline B {}^0_s\to \overline D^0\pi^0}\\
A^0_{\overline B {}^0_s\to \overline D^-_s K^+}\\
A^0_{\overline B {}^0_s\to \overline D^0\eta_8}\\
A^0_{\overline B {}^0_s\to \overline D^0\eta_1}
\end{array}
\right).
 \label{eq:FSIBsDbarP}
\end{eqnarray}

In these expressions, the square root of the rescattering
$S$-matrix are denoted as ${\cal S}^{1/2}_i=(1+i{\cal
T}_i)^{1/2}=1+i{\cal T}^\prime_i$, with
\begin{eqnarray}
 {\cal T}_1 &=& \left(
\begin{array}{cc}
r_0 &r_e\\
r_e &r_0
\end{array}
\right), \nonumber\\
{\cal T}_2&=&\left(
\begin{array}{ccccc}
r_0+r_a
       &\frac{r_a-r_e}{\sqrt2}
       &r_a
       &\frac{r_a+r_e}{\sqrt6}
       &\frac{\bar r_a+\bar r_e}{\sqrt3}
       \\
\frac{r_a-r_e}{\sqrt2}
       &r_0+\frac{r_a+r_e}{2}
       &\frac{r_a}{\sqrt2}
       &\frac{r_a+r_e}{2\sqrt3}
       &\frac{\bar r_a+\bar r_e}{\sqrt6}
       \\
r_a
       &\frac{r_a}{\sqrt2}
       &r_0+r_a
       &\frac{r_a-2 r_e}{\sqrt6}
       &\frac{\bar r_a+\bar r_e}{\sqrt3}
       \\
\frac{r_a+r_e}{\sqrt6}
       &\frac{r_a+r_e}{2\sqrt3}
       &\frac{r_a-2 r_e}{\sqrt6}
       &r_0+\frac{r_a+r_e}{6}
       &\frac{\bar r_a+\bar r_e}{3\sqrt2}
       \\
\frac{\bar r_a+\bar r_e}{\sqrt3}
       &\frac{\bar r_a+\bar r_e}{\sqrt6}
       &\frac{\bar r_a+\bar r_e}{\sqrt3}
       &\frac{\bar r_a+\bar r_e}{3\sqrt2}
       &\tilde r_0+\frac{\tilde r_a+\tilde r_e}{3}
\end{array}
\right), \nonumber\\
{\cal T}_3&=&\left(
\begin{array}{ccccc}
r_0+r_a
       &r_a
       &\frac{r_e}{\sqrt2}
       &\frac{r_e-2 r_a}{\sqrt6}
       &\frac{\bar r_e+\bar r_a}{\sqrt3}
       \\
r_a
       &r_0+r_a
       &-\frac{r_e}{\sqrt2}
       &\frac{r_e-2 r_a}{\sqrt6}
       &\frac{\bar r_a+\bar r_e}{\sqrt3}
       \\
\frac{r_e}{\sqrt2}
       &-\frac{r_e}{\sqrt2}
       &r_0
       &0
       &0
       \\
\frac{r_e-2 r_a}{\sqrt6}
       &\frac{r_e-2 r_a}{\sqrt6}
       &0
       &r_0+\frac{2}{3}(r_a+r_e)
       &-\frac{\sqrt2}{3}(\bar r_a+\bar r_e)
       \\
\frac{\bar r_e+\bar r_a}{\sqrt3}
       &\frac{\bar r_a+\bar r_e}{\sqrt3}
       &0
       &-\frac{\sqrt2}{3}(\bar r_a+\bar r_e)
       &\tilde r_0+\frac{\tilde r_a+\tilde r_e}{3}
\end{array}
\right), \label{eq:T}
\end{eqnarray}
where $r_e$, $r_a$ and $r_0$ are charge exchange, annihilation,
singlet exchange rescattering parameters~\cite{Chua:2001br}, while
$\bar r_i$ and $\tilde r_i$ are those for $D\Pi({\bf
8})\leftrightarrow D^0\eta_1$ and $D^0\eta_1\leftrightarrow
D^0\eta_1$ scattering, respectively~\cite{Chua:2005dt}. SU(3)
symmetry requires that ${\cal T}^\prime_i$ has the same structure
as ${\cal T}_i$. Hence the ${\cal T}'_i$ is basically ${\cal
T}_i$, but with $r_j$, $\bar r_j$ and $\tilde r_j$ replaced by
$r_j^\prime$, $\bar r'_j$ and $\tilde r'_j$, respectively. We note
that some of the above formulas were already reported in
\cite{Chua:2001br,Chua:2005dt}, while all formulas for the second
and fourth cases, and some for the third case, are new. We have
used charge conjugation invariance and SU(3) symmetry of the
strong interactions, hence the $r^{(\prime)}_i$, $\bar
r^{(\prime)}_i$ and $\tilde r^{(\prime)}_i$ coefficients in ${\cal
T}^{(\prime)}_i$ of $\overline B_{u,d,s}\to \overline DP$
rescattering amplitudes are identical to those in ${\cal
T}^{(\prime)}_{i}$ of $\overline B_{u,d,s}\to D P$ rescattering
amplitudes.

Using SU(3) symmetry and ${\cal S}^\dagger {\cal S}=1$, the
rescattering parameters are given by~\cite{Chua:2005dt}
 \begin{eqnarray}
 (1+i r_0)&=&\frac{1}{2}(1+e^{2i\delta}),
 \nonumber\\
 i r_e &=&\frac{1}{2}(1-e^{2i\delta}),
 \nonumber\\
 i r_a &=&\frac{1}{8}(3 {\cal U}_{11}-2 e^{2i\delta}-1),
 \nonumber\\
 i(\bar r_a+\bar r_e)&=&\frac{3}{2\sqrt2} {\cal U}_{12},
 \nonumber\\
 i(\tilde r_0+\frac{\tilde r_a+\tilde r_e}{3})&=&{\cal U}_{22}-1,
 \label{eq:solution}
 \end{eqnarray}
where
\begin{equation}
{\cal U}={\cal U}^T=\left(
\begin{array}{cc}
\cos\tau
       &\sin\tau
       \\
-\sin\tau
       &\cos\tau
\end{array}
\right)
\left(
\begin{array}{cc}
e^{2i\theta}
       &0
       \\
0
       &e^{2i\sigma}
\end{array}
\right)
\left(
\begin{array}{cc}
\cos\tau
       &-\sin\tau
       \\
\sin\tau
       &\cos\tau
\end{array}
\right), \label{eq:U}
\end{equation}
and we have set the overall phase factor ($1+ir_0+ir_e$) in ${\cal
S}$ to unity. This phase convention is equivalent to choosing the
$A_{\overline B {}^0\to D^0\pi^-}$ amplitude to be real.
The $r'_i$, $\bar r'_i$ and $\tilde r'_i$ in ${\cal S}^{1/2}$ can
be obtained by using the above formulas with phases
($\delta,\,\theta,\,\sigma$) reduced by half.
We need three phases and one mixing angle to specify FSI effects
in $DP$ and $\overline D P$ rescattering. The interpretation of
these phases and mixing angle in term of SU(3) decomposition can
be found in~\cite{Chua:2005dt}.

To use the FSI formulas, we need to specify $A^0$. We use naive
factorization amplitudes $A^f$ for $A^0$ to avoid double counting
of FSI effects~\cite{Chua:2001br,Chua:2005dt}. And the explicit
forms of $A^0$ are given in Appendix A.
We stress that, in our quasi-elastic rescattering approach, SU(3)
symmetry is applied only in the $D\Pi \to D\Pi$ rescattering
matrix, which should hold for $m_B$ scale. Certain amount of SU(3)
breaking effects which have to do with meson formation are
included in the factorization amplitudes via decay constants and
form factors.

\begin{table}[t!]
\caption{ \label{tab:table-br} Branching ratios of various
$\overline B\to DP$ and $D\overline K$ modes in $10^{-4}$ units.
The second column is the experimental data~\cite{HFAGc,PDG}, which
is taken as input. The naive factorization model results are given
in third column. Fitting the experimental data shown in the first
column with quasi-elastic FSI (fit parameters as given in Table
II), we obtain the FSI fit results given in the last column. The
factorization results are recovered by setting FSI phases in
Table~II to zero.
}
\begin{ruledtabular}
\begin{tabular}{lccc}
 Mode
      &$\mathcal{B}^{\rm exp}$ ($10^{-4}$)
      &$\mathcal{B}^{\rm fac}$ ($10^{-4}$)
      &$\mathcal{B}^{\rm FSI}$ ($10^{-4}$)
      \\
\hline
 $B^-\to D^0\pi^-$
        & $48.4\pm1.5$
        & $48.4^{+5.2}_{-4.2}$
        & $48.4^{+0.8}_{-0.8}$
        \\
\hline $\overline B {}^0\to D^+\pi^-$
        & $26.8\pm1.3$
        & $31.9^{+1.7}_{-1.8}$
        & $26.9^{+1.0}_{-1.0}$
        \\
$\overline B {}^0\to D^0\pi^0$
        & $2.61\pm 0.24$
        & $0.57^{+0.25}_{-0.14}$
        & $2.42^{+0.19}_{-0.16}$
        \\
$\overline B {}^0\to D^+_s K^-$
        & $0.28\pm0.05$
        & 0
        & $0.26\pm0.03$
        \\
$\overline B {}^0\to D^0\eta$
        & $2.02\pm0.35$
        & $0.33^{+0.14}_{-0.08}$
        & $2.06^{+0.30}_{-0.29}$
        \\
$\overline B {}^0\to D^0\eta'$
        & $ 1.25\pm 0.23 $
        & $0.20^{+0.09}_{-0.05}$
        & $1.27^{+0.21}_{-0.19}$
        \\
\hline $B^-\to D^0 K^-$
        & $4.02\pm0.21$
        & $4.01^{+0.49}_{-0.38}$
        & $4.01^{+0.07}_{-0.09}$
        \\
\hline
 $\overline B {}^0\to D^+K^-$
        & $2.04\pm 0.57$
        & $2.43\pm 0.13$
        & $1.97\pm0.07$
        \\
$\overline B {}^0\to D^0 \overline K{}^0$
        & $0.52\pm0.07$
        & $0.14^{+0.06}_{-0.04}$
        & $0.60^{+0.03}_{-0.04}$
        \\
\end{tabular}
\end{ruledtabular}
\end{table}

\section{\label{sec:num} Results}

In our numerical study, masses and lifetimes are taken from the
Particle Data Group (PDG)~\cite{PDG}, and $B$ to charm meson decay
branching ratios are taken from~\cite{HFAGc,PDG}. We fix $V_{ud}=
0.97419$, $V_{us}= 0.22568$, $V_{cb}= 0.04166$, $V_{cs}=0.997334$,
$|V_{ub}|=3.624\times 10^{-3}$~\cite{CKMfitter}, and use the decay
constants $f_\pi=$ 131 MeV, $f_{K}=$ 156 MeV~\cite{PDG} and
$f_{D_{(s)}}=$ 200 (230) MeV.

We have six parameters to describe the processes with rescattering
from factorization amplitudes: the two effective Wilson
coefficients $a^{\rm eff}_1$ and $a^{\rm eff}_2$, the three
rescattering phases $\delta$, $\theta$ and $\sigma$, and one
mixing angle $\tau$ in ${\cal S}^{1/2}$. These parameters are
fitted with rates of nine $\overline B$ decay to $C = 1$, $S =0,\;
-1$ modes, namely $\overline B \to D^+\pi^-$, $D^0\pi^-$,
$D^0\pi^0$, $D^0\eta$, $D^0\eta'$, $D^+_s K^-$, $D^0 K^-$, $D^+
K^-$ and $D^0\overline K {}^0$ decays, given in Table~I. The
fitted FSI parameters are listed in Table II. We then use the
extracted parameters to predict nineteen $\overline B_{s}\to DP$
(Table III) and $\overline B_{u,d,s}\to \overline D P$ (Table~IV)
decays. Predictions on the ratios of $\overline B\to \overline D\,
P$ and $\overline B\to D P$ amplitudes are also given.

\begin{table}[t!]
\caption{ \label{tab:airi} Fit parameters in the SU(3) FSI
picture, where results are from using $\overline B \to D^0\pi^-$,
$D^+\pi^-$, $D^0\pi^0$, $D^0\eta$, $D^0\eta'$, $D^+_s K^-$, $D^0
K^-$, $D^+ K^-$ and $D^0 \overline K {}^0$ decay rates
(Table~\ref{tab:table-br}) as fit input. There is a two fold
ambiguity (the overall sign of phases) in the solutions. The SU(3)
phases and mixing angle are reexpressed in terms of the
rescattering parameters $r_i^\prime$,
$\bar r_i^\prime$, $\tilde r_i^\prime$. 
}
\begin{ruledtabular}
\begin{tabular}{lclc}
      parameter
      & result
      &parameter
      & result
      \\
 \hline
 $\chi^2_{\rm min}$
 &1.92
 &$\chi^2_{\rm min}/{\rm d.o.f.}$
 &0.64
 \\
 \hline
 $a^{\rm eff}_1$
     &$0.90 \pm 0.02$
     &$a^{\rm eff}_2$
     &$0.23^{+0.03}_{-0.02}$
     \\
 $\delta$
     &$\pm(52.9^{+1.9}_{-2.0})^\circ$
     &$\theta$
     &$\pm(17.8^{+3.0}_{-2.9})^\circ$
     \\
 $\sigma$
     &$\mp(87.7^{+27.7}_{-27.3})^\circ$
     &$\tau$
     &$(9.2^{+5.2}_{-2.7})^\circ$
     \\
\hline
 $1+i r'_0$
        &$(0.80\pm0.01)\pm (0.40\pm0.01)i$
 &$i r'_e$
        &$(0.20\pm0.01)\mp (0.40\pm0.01)i$
        \\
 $i r'_a$
        &$(0.07\pm0.01)\mp(0.10\pm0.01)i$
 &$i(\bar r'_a+\bar r'_e)$
        &$(-0.15^{+0.04}_{-0.03})\mp (0.22^{+0.08}_{-0.09})i$
        \\
 $1+i \tilde r'_0+i\frac{\tilde r'_e+\tilde r'_a}{3}$
        &$(0.06\pm0.46)\mp(0.97^{+0.17}_{-0.01})i$
\end{tabular}
\end{ruledtabular}
\end{table}

The errors of the fitted parameters given in Table~\ref{tab:airi}
are propagated from the experimental errors by requiring
$\chi^2\leq\chi^2_{\rm min.}+1$. The fitted values of these
parameters are similar to those in our previous
analysis~\cite{Chua:2005dt}~\footnote{
 We found and corrected a numerical error in our previous analysis,
 resulting in the value of $\sigma$ taking opposite sign.}. There is a two fold
ambiguity (the overall sign of the phases) in the solutions. We
obtain $\chi^2_{\rm min}/{\rm d.o.f.}=0.64$ indicating a good fit
to these modes.
The effective Wilson coefficients $a^{\rm eff}_{1,2}$ are close to
expectation~\cite{Neubert:1997uc,Cheng:1999kd}. From
$|r'_e|>|r'_a|$ we infer that exchange rescattering is dominant
over annihilation rescattering.

We show in the fourth column of Table~\ref{tab:table-br} the fit
output for the nine fitted $\overline B\to DP$ and $D\overline K$
modes. These fitted branching ratios (in units of $10^{-4}$)
should be compared with data and naive factorization results given
in the second and third columns. The FSI results reproduce the
data quite well, as it should. The errors for the FSI results are
from data only. The factorization results can be recovered by
using the parameters of Table II but with FSI phases set to zero.
Note that unitarity is implied automatically, i.e. sum of rates
within coupled modes are unchanged by FSI.

\begin{table}[t!]
\caption{ \label{tab:table-br-prediction} The predictions on
branching ratios of various $\overline B_s\to DP$ modes in
$10^{-4}$ and $10^{-5}$ units, respectively. The errors for the
FSI results are from $\overline B\to DP$ data only.
}
\begin{ruledtabular}
\begin{tabular}{lccc}
 Mode
      &$\mathcal{B}^{\rm exp}$
      &$\mathcal{B}^{\rm fac}$ ($10^{-4}$)
      &$\mathcal{B}^{\rm FSI}$ ($10^{-4}$)
      \\
 \hline
$\overline B {}^0_s\to D^+_s \pi^-$
        & $30\pm7$
        & $30.5^{+1.6}_{-1.7}$
        & $24.9^{+0.8}_{-0.9}$
        \\
$\overline B {}^0_s\to D^0 K{}^0$
        & --
        & $2.2^{+1.0}_{-0.6}$
        & $7.9\pm{0.5}$
        \\
\hline
 Mode
      &$\mathcal{B}^{\rm exp}$ ($10^{-5}$)
      &$\mathcal{B}^{\rm fac}$ ($10^{-5}$)
      &$\mathcal{B}^{\rm FSI}$ ($10^{-5}$)
       \\
       \hline
$\overline B {}^0_s\to D^+\pi^-$
        & --
        & 0
        & $0.16^{+0.03}_{-0.02}$
        \\
$\overline B {}^0_s\to D^0\pi^0$
        & --
        & 0
        & $0.08\pm0.01$
        \\
$\overline B {}^0_s\to D^+_s K^-$
        & --
        & $23.2^{+1.2}_{-1.3}$
        & $19.5^{+0.7}_{-0.7}$
        \\
$\overline B {}^0_s\to D^0\eta$
        & --
        & $0.6^{+0.3}_{-0.2}$
        & $2.9^{+0.5}_{-0.5}$
        \\
$\overline B {}^0_s\to D^0\eta'$
        & --
        & $0.9^{+0.4}_{-0.2}$
        & $2.0\pm0.6$
\end{tabular}
\end{ruledtabular}
\end{table}


Our main interest here is the color-suppressed $B_s$ decays. The
predicted branching ratios of various $\overline B_s\to DP$ modes
with $C=+1,\;S=0,\;+1$ final states are shown in
Table~\ref{tab:table-br-prediction}, where the second column gives
naive factorization results and the third column gives the FSI
results. Again, the factorization results are recovered by using
the same parameters of Table II but with FSI phases set to zero,
and the errors for the FSI results are from $\overline B\to DP$
data only.
Analogous to $\overline B^0 \to D^0\pi^0$ enhancement being fed
from $\overline B^0 \to D^+\pi^-$ rescattering, it is interesting
to note that $\overline B^0_s \to D_s^+\pi^-$ with FSI
rescattering to $D^0 K{}^0$, brings $\overline B {}^0_s\to D^0
K{}^0$ rate to the $10^{-3}$ level, which can be measured soon.
This is helped by the absence of annihilation rescattering. The
$\overline B_s\to D^0\eta,\,D^0\eta'$ modes are the direct analogs
of $\overline B^0 \to D^0\pi^0$. One can see that their rates are
brought up to levels similar to $\overline B^0 \to D^0\pi^0$.
Rescattering slightly reduces the $\overline B_s\to D_s^+ K^-$ and
$B_s\to D_s^+ \pi^-$ rates. The $D^+_s K$ mode will be used to
extract $\gamma/\phi_3$ at LHCb~\cite{ADK}, while the
$D^0\eta,\,D^0\eta'$ modes could also be
useful~\cite{Chua:2005fj}.

\begin{table}[t!]
\caption{ \label{tab:table-br1} Predictions for $\overline
B_{u,d,s}\to \overline D P$ rates. Experimental results and
limits~\cite{PDG,DbarK0bar} are shown in the second column, and
naive factorization and FSI results are given in the third and
fourth columns.
}
\begin{ruledtabular}
\begin{tabular}{lccc}
 Mode
      &$\mathcal{B}^{\rm exp}$ ($10^{-5}$)
      &$\mathcal{B}^{\rm fac}$ ($10^{-5}$)
      &$\mathcal{B}^{\rm FSI}$ ($10^{-5}$)
      \\
\hline
 $B^-\to \overline D {}^0 K^-$
        & --
        &$0.2\pm0.1$
        &$0.3^{+0.1}_{-0.3}$
        \\
 $B^-\to D^- \overline K{}^0$
        & $<0.5$
        & 0
        &$0.03\pm0.01$
        \\
$B^-\to D_s^-\pi^0$
        & $<20$
        & $0.9\pm0.1$
        & $0.8\pm0.0$
        \\
$B^-\to D_s^-\eta$
        & $<50$
        & $0.5\pm0.0$
        & $0.4\pm0.1$
        \\
$B^-\to D_s^-\eta'$
        & --
        & $0.3\pm0.0$
        & $0.5\pm0.1$
        \\
        \hline
$\overline B {}^0\to D^-_s \pi^+$
        & $1.4\pm0.3$
        & $1.7\pm0.1$
        & $1.4^{+0.0}_{-0.1}$
        \\
$\overline B {}^0\to \overline D {}^0 K{}^0$
        & --
        & $0.2\pm0.1$
        & $0.5\pm0.0$
        \\
       \hline
$\overline B {}^0_s\to D^-\pi^+$
        & --
        & 0
        & $0.02\pm0.00$
        \\
$\overline B {}^0_s\to \overline D{}^0\pi^0$
        & --
        & 0
        & $0.01\pm0.00$
        \\
$\overline B {}^0_s\to D^-_s K^+$
        & --
        & $2.3\pm0.1$
        & $2.0\pm{0.1}$
        \\
$\overline B {}^0_s\to \overline D {}^0\eta$
        & --
        & $0.06^{+0.03}_{-0.02}$
        & $0.3^{+0.0}_{-0.1}$
        \\
$\overline B {}^0_s\to \overline D {}^0\eta'$
        & --
        & $0.09^{+0.04}_{-0.02}$
        & $0.2\pm0.1$
\end{tabular}
\end{ruledtabular}
\end{table}

$\overline B_{u,d,s}\to \overline D P$ decays are $V_{ub}$
suppressed, and mostly not measured yet, except $D_s^-\pi^+$. The
quasi-elastic FSI formalism allows us to make predictions even for
such modes that are color-suppressed. Predictions for $\overline
B_{u,d,s}\to \overline D P$ decays are shown in
Table~\ref{tab:table-br1}, where again, experimental results and
limits~\cite{PDG,DbarK0bar} are shown in the second column, and
the third and fourth columns are naive factorization and FSI
results, respectively. Same comments on FSI parameters apply.
Agreement of the only observed $\overline B {}^0\to D^-_s \pi^+$
mode with theoretical prediction is improved by including FSI
effects. $\overline B_{s}\to D_s^- K^+$ is slightly reduced, but
overall, the redistribution of decay rates by rescattering is not
very significant.

Combining $\overline B_s\to D^+_s K^-$ and $\overline B_s\to D^-_s
K^+$, we can compare our predicted ratio
 \begin{eqnarray}
 R\equiv\frac{{\cal B}(\overline B_s\to D^\pm_s K^\mp)}{{\cal B}(\overline B_s\to
 D^+_s\pi^-)}=0.090\pm0.002,
 \end{eqnarray}
with the recent experimental result of
$R=0.107\pm0.019\pm0.008$~\cite{CDF} from CDF. The agreement is
reasonable. In fact, the larger rescattering of $\overline B_s\to
 D^+_s\pi^-$ to $\overline B_s\to
 D^0K^0$ has helped enhance the ratio from the lower factorization
 value.

\begin{table}[t!]
\caption{ \label{tab:table-br-comparison} Comparison of
predictions for branching ratios of various $\overline B_s\to DP$
modes to other approaches.}
\begin{ruledtabular}
\begin{tabular}{lccc}
 $\mathcal{B}$ ($10^{-4}$)
      &This work 
      &CF~\cite{CF} 
      &CS~\cite{CS} 
      \\
 \hline
$\overline B {}^0_s\to D^+_s \pi^-$
        & $24.9^{+0.8}_{-0.9}$
        & $29\pm 6$
        & $22\pm 1$
        \\
$\overline B {}^0_s\to D^0 K{}^0$
        & $7.9\pm0.5$
        & $8.1\pm1.8$
        & $5.3\pm0.3$
        \\
\hline
 $\mathcal{B}$ ($10^{-5}$) 
      & This work
      & CF~\cite{CF}
      & CS~\cite{CS}
       \\
       \hline
$\overline B {}^0_s\to D^+\pi^-$
        & $0.16^{+0.03}_{-0.02}$
        & $0.20\pm0.06$
        & $0.14\pm0.03$
        \\
$\overline B {}^0_s\to D^0\pi^0$
        & $0.08\pm0.01$
        & $0.10\pm 0.03$
        & $0.07\pm0.01$
        \\
$\overline B {}^0_s\to D^+_s K^-$
        & $19.5\pm{0.7}$
        & $18\pm 3$
        & $20\pm1$
        \\
$\overline B {}^0_s\to D^0\eta$
        & $2.9\pm{0.5}$
        & $2.1\pm 1.2$
        & $1.4\pm0.1$
        \\
$\overline B {}^0_s\to D^0\eta'$
        & $2.1\pm0.6$
        & $0.98\pm0.76$
        & $2.9\pm0.2$
\end{tabular}
\end{ruledtabular}
\end{table}


In Table~\ref{tab:table-br-comparison}, we compare our predictions
for various $\overline B_s\to DP$ rates with results obtained in
other approaches~\cite{CF,CS} that differ in the application of
SU(3) symmetry. Most of our results agree with others. For modes
with $\eta^{(\prime)}$, our results are closer to those in
\cite{CF} obtained using earlier data. In both approaches, U(3)
symmetry is not imposed and $D\eta_1$ is treated as an independent
component.
Although predictions on $\overline B_s\to DP$ rates are similar in
all three works, it should be note that there is a major
difference between ours and the other two's approaches. In this
work, the information obtained in $\overline B_{u,d}\to DP$
rescattering from data is used to predict not only $\overline
B_s\to DP$ decays [via SU(3) symmetry], but also $\overline
B_{u,d,s}\to \overline D P$ decays [through charge conjugation
invariance of the $S$-matrix]. SU(3) symmetry itself is not
sufficient to relate $\overline B_{u,d,s}\to D P$ and $\overline
B_{u,d,s}\to \overline D P$ amplitudes. Hence, in the two other
works, which employed solely SU(3) symmetry to decay amplitudes,
no prediction on $\overline B_{u,d,s}\to \overline D P$ decays
were given by analyzing $\overline B_{u,d}\to DP$ data.

In Table~\ref{tab:rB}, $r_{B(B_S)}$, $\delta_{B(B_S)}$ for various
modes are predicted and compared with data, where the amplitude
ratio $r_{B_{(s)}}$ and the strong phase difference
$\delta_{B_{(s)}}$ are defined as
 \begin{equation}
 r_{B_{(s)}}(DP)=\left|
 \frac{A(\overline B_{(s)}\to \overline D\, \overline P)}
      {A(\overline B_{(s)}\to D P)}\right|,
 \qquad
 \delta_{B_{(s)}}(DP)=\arg\left[
 \frac{e^{i\phi_3} A(\overline B_{(s)}\to \overline D\, \overline P)}
  {A(\overline B_{(s)}\to D P)}\right].
 \label{eq:rB}
 \end{equation}
The weak phase $\phi_3$ is removed from $A(\overline B_{(s)}\to
\overline D \bar P)$ in defining $\delta_{B(B_s)}$. Except
$r_B(D^0 K^0)$ the effects from final state interaction are mild.
We see that our $r_B(D^0 K^-)$ and $\delta_B(D^0 K^-)$ agree with
the Dalitz analysis results of Belle and BaBar. Our $r_B(D^0 K^-)$
is also in agreement with the fit from UT$_{fit}$ group obtained
by using all three methods of GLW, ADS and $DK$ Dalitz
analysis~\cite{UTfit}.

\begin{table}[t!]
\caption{ \label{tab:rB} Naive factorization and FSI results on
$r_{B_{(s)}}$, $\delta_{B_{(s)}}$ with
$|V_{ub}|=3.67\times10^{-3}$, and compared to the experimental
results~\cite{UTfit,HFAG}. The errors for the FSI results
are from $DP$ data only.} 
\begin{ruledtabular}
\begin{tabular}{lccc}
      &Expt
      &fac
      &FSI
      \\
\hline
 $r_B(D^0 K^-)$
        &$0.16\pm0.05\pm0.01\pm0.05$ (Belle)
        &$0.07\pm0.01$
        &$0.09\pm0.01$
        \\
        &$<0.14$ $(1\sigma)$ (BaBar)
        \\
        &$0.071\pm0.024$ (UT$_{fit}$)
        \\
 $\delta_B(D^0 K^-)$
        &$(146^{+19}_{-20}\pm3\pm23)^\circ$ (Belle)
        &$180^\circ$
        &$180^\circ\mp(39.4^{+5.4}_{-6.4})^\circ$
        \\
        &$(118\pm 63\pm 19\pm36)^\circ$ (BaBar)
        &
        &
        \\
        \hline
 $r_B(D^0 K^0)$
        & --
        &$0.38\pm0.00$
        &$0.29\pm0.01$
        \\
 $\delta_B(D^0 K^0)$
        & --
        & $180^\circ$
        & $180^\circ\pm(8.6^{+0.7}_{-0.5})^\circ$
        \\
        \hline
 $r_{B_s}(D_s^+ K^-)$
        & --
        &$0.38\pm0.00$
        &$0.38\pm0.00$
        \\
 $\delta_{B_s}(D_s^+ K^-)$
        & --
        & $180^\circ$
        & $180^\circ\pm(0.1\pm0.0)^\circ$
        \\
        \hline
 $r_{B_s}(D^0 \eta)$
        & --
        &$0.38\pm0.00$
        &$0.38\pm0.00$
        \\
 $\delta_{B_s}(D^0 \eta)$
        & --
        & $180^\circ$
        & $180^\circ\mp(0.6\pm 0.0)^\circ$
        \\
        \hline
 $r_{B_s}(D^0 \eta')$
        & --
        &$0.38\pm0.00$
        &$0.38\pm0.00$
        \\
 $\delta_{B_s}(D^0 \eta')$
        & --
        & $180^\circ$
        & $180^\circ\mp(0.2\pm 0.1)^\circ$
        \\
\end{tabular}
\end{ruledtabular}
\end{table}

\section{Conclusion}

We study quasi-elastic rescattering effects in $\overline
B_{u,d,s}\to DP$, $\overline D P$ modes. The updated data for nine
$\overline B_{u,d} \to D P$ modes are used to extract
$a_{1,2}^{\rm eff}$ and four rescattering parameters.
We find the effective Wilson coefficients $a^{\rm eff}_1\simeq
0.90$, $a^{\rm eff}_2\simeq 0.23$, the strong phases $\delta\simeq
54^\circ$, $\theta\simeq18^\circ$, $\sigma\simeq -88^\circ$ and
mixing angle $\tau\simeq 9^\circ$. Since strong interaction
respects SU(3) symmetry and charge conjugation symmetry, the
formalism can be used to predict $\overline B_s\to D P$ and
$\overline B_{u,d,s}\to \overline D P$ rates and $r_{B(B_s)}$,
without refers to any experimental information on $\overline
B_{u,d,s}\to \overline D P$ decays, which is limited by the
smallness of the corresponding decay rates.

Our results are summarized as following: (a) The $\overline B
{}^0_s\to D^0 K{}^0,\, D^0\eta,\,D^0\eta'$ rates are enhanced in
the presence of FSI. In particular, the $\overline B {}^0_s\to D^0
K{}^0$ rate is close to $10^{-3}$ level and can be measured soon.
(b) The predicted $\overline B_s\to D_s^+\pi^-$ rate and the ratio
of ${\cal B}(\overline B_s\to D^\pm_s K^\mp)/{\cal B}(\overline
B_s\to D^+_s\pi^-)$ is in better agreement with experimental
results. (c) Except the $\overline B^0\to D^0 K^0$ mode, the FSI
effects on $r_{B(B_S)}$ are mild. (d) The predicted $r_B(D^0 K^-)$
agree with data and the fit from the UTfit collaboration, while
$\delta_B(D^0 K^-)$ agree with data.

\begin{acknowledgments}
 We would like to thank Alexey Drutskoy for useful discussion.
This work is supported in part by the National Science Council of
R.O.C. under Grants NSC-95-2112-M-033-MY2 and
NSC96-2112-M-002-022.
\end{acknowledgments}

\appendix

\section{Explicit expression of $A^0$}

As mentioned in the text, we use naive factorization amplitudes
$A^f$ for $A^0$ to avoid double counting of FSI effects. For each
final state, we have
 \begin{eqnarray}
 A^f_{B^-\to D^0\pi^-}&=&V_{cb} V_{ud}^*(T_f+C_f),
 \qquad
 A^f_{\overline B {}^0\to D^+\pi^-}=V_{cb} V_{ud}^*(T_f+E_f),
 \nonumber \\
 A^f_{\overline B {}^0\to D^0\pi^0}&=&\frac{V_{cb} V_{ud}^*}{\sqrt2}(-C_f+E_f),
 \quad
 A^f_{\overline B {}^0\to D^+_s K^-}=V_{cb} V_{ud}^*~E_f,
 \nonumber \\
 A^f_{\overline B {}^0\to D^0\eta_8}&=&\frac{V_{cb} V_{ud}^*}{\sqrt6}(C_f+E_f),
 \qquad
 A^f_{\overline B {}^0\to D^0\eta_1}=\frac{V_{cb} V_{ud}^*}{\sqrt3}(C_f+E_f),
 \nonumber\\
 A_{B^-\to D^0 K^-}&=&V_{cb} V_{us}^*(T_f+C_f),
 \,\,\,\qquad
 A_{\overline B {}^0\to D^+ K^-}=V_{cb} V_{us}^*~T_f,
 \nonumber\\
 A_{\overline B {}^0\to D^0 \overline K^0}&=&V_{cb} V_{us}^*~C_f,
 \quad\qquad\qquad
 A^f_{\overline B {}^0_s\to D^+_s \pi^-}=V_{cb} V_{ud}^* T_f,
 \nonumber\\
 A^f_{\overline B {}^0_s\to D^0 K^0}&=&V_{cb} V_{ud}^*~C_f,
 \quad\qquad\qquad
 A^f_{\overline B {}^0_s\to D^+\pi^-}=V_{cb} V_{us}^*~E_f,
 \nonumber\\
 A^f_{\overline B {}^0_s\to D^0\pi^0}&=&\frac{V_{cb} V_{us}^*}{\sqrt2}~E_f,
 \quad\qquad\qquad
 A^f_{\overline B {}^0_s\to D^0\eta_8}=\frac{V_{cb} V_{us}^*}{\sqrt6}~(-2C_f+E_f),
 \nonumber\\
 A^f_{\overline B {}^0_s\to D^0\eta_1}&=&\frac{V_{cb} V_{us}^*}{\sqrt3}~(C_f+E_f),
 \,\,\,\quad
 A^f_{\overline B {}^0_s\to D^+_s K^-}=V_{cb} V_{us}^*~(T_f+E_f),
 \label{eq:Atop}
\end{eqnarray}
where the super- and subscripts $f$ indicate naive factorization
amplitude, and
\begin{eqnarray}
T_f&=&{G_F\over\sqrt2} \, a^{\rm eff}_1 \, (m_B^2-m_D^2)
           f_P F_0^{BD}(m^2_P),
\nonumber\\
C_f&=&{G_F\over\sqrt2} \, a^{\rm eff}_2 \,
                (m_B^2-m_P^2) f_D F_0^{BP}(m^2_D),
\nonumber\\
E_f&=&{G_F\over\sqrt2} \, a^{\rm eff}_2 \,
                (m_D^2-m_P^2) f_B F_0^{0\to DP}(m^2_B).
\end{eqnarray}
$F_0^{BD(BP)}$ is the $\overline B_{u,d,s} \to D_{u,d,s}(P)$
transition form factor and $F_0^{0\to DP}$ is the vacuum to $DP$
(time-like) form factor.

For $\overline B_{u,d,s}\to\overline D_{u,d,s} P$ decays, we have,
 \begin{eqnarray}
 A^f_{B^-\to\overline D {}^0 K^-}&=&V_{ub} V_{cs}^*~(c_f+a_f),
 \qquad
 A^f_{B^-\to D^- \overline K{}^0}=V_{ub} V_{cd}^*~a_f,
 \nonumber \\
 A^f_{B^-\to\overline D{}^0\eta_8}&=&\frac{V_{ub} V_{cs}^*}{\sqrt6} (t_f-2 a_f),
 \,\,\,\quad
 A^f_{B^-\to\overline D{}^0\eta_1}=\frac{V_{ub} V_{cd}^*}{\sqrt3}(t_f+a_f),
 \nonumber\\
 A^f_{B^-\to D^-_s\pi^0}&=&\frac{V_{ub} V_{cs}^*}{\sqrt2} t_f,
 \quad\qquad\qquad
 A_{\overline B {}^0\to D_s^- \pi^+}=V_{ub} V_{cs}^*~t_f,
 \nonumber\\
 A_{\overline B {}^0\to \overline D {}^0 \overline K{}^0}&=&V_{ub} V_{cs}^*~c_f,
 \quad\qquad\qquad
 A^f_{\overline B {}^0_s\to D^-\pi^+}=V_{ub} V_{cs}^*~e_f,
 \nonumber \\
 A^f_{\overline B {}^0_s\to \overline D {}^0\pi^0}&=&\frac{V_{ub} V_{cs}^*}{\sqrt2}~e_f,
 \quad\qquad\qquad
 A^f_{\overline B {}^0_s\to \overline D {}^0\eta_8}=\frac{V_{ub} V_{cs}^*}{\sqrt6}~(-2c_f+e_f),
 \nonumber \\
 A^f_{\overline B {}^0_s\to \overline D {}^0\eta_1}&=&\frac{V_{ub} V_{cs}^*}{\sqrt3}~(c_f+e_f),
 \qquad
 A^f_{\overline B {}^0_s\to D^-_s K^+}=V_{ub} V_{cs}^*~(t_f+e_f),
 \label{eq:Atop}
\end{eqnarray}
where, as before, the super- and subscripts $f$ indicate naive
factorization amplitude, and
\begin{eqnarray}
t_f&=&{G_F\over\sqrt2} \, a^{\rm eff}_1 \, (m_B^2-m_P^2)
           f_D F_0^{BP}(m^2_D),
\nonumber\\
c_f&=&{G_F\over\sqrt2} \, a^{\rm eff}_2 \,
                (m_B^2-m_P^2) f_D F_0^{BP}(m^2_D),
\nonumber\\
e_f&=&{G_F\over\sqrt2} \, a^{\rm eff}_2 \,
                (m_P^2-m_D^2) f_B F_0^{0\to PD}(m^2_B).
\end{eqnarray}
Note that we have $(t_f,\,\,e_f)=(T_f,\,E_f)$ with $D$ and $P$
interchanged and $c_f=C_f$ (without the interchange of $D$ and
$P$).

\begin{table}[t!]
\caption{ Form factors taken from~\cite{LF,MS}. For
$B_{(s)}\to\eta^{\prime}$ form factors the mixing angle and
Clebsch-Gordan coefficients are included [see
Eq.~(\ref{eq:FBtoeta})].
 \label{tab:formfactor}
}
\begin{ruledtabular}
\begin{tabular}{lclc}
Form factor
 & value
 &Form factor
 & value
 \\
  \hline
 $F_0^{B\pi}(m_{D,D_s}^2)$
  & 0.28
  &$F_0^{B_{(s)}D_{(s)}}(m_{\pi,K}^2)$
  & 0.67
  \\
$F_0^{B\eta}(m_{D,D_s}^2)$
  & 0.15
  &$F_0^{B\eta^\prime}(m_{D,D_S}^2)$
  &0.13 \\
$F^{BK}_0(m^2_{D,D_s})$
  &0.43
  & $F^{BsK}_0(m^2_{D})$
  & 0.40
  \\
$F_0^{B_s\eta}(m_{D}^2)$
  & $-0.29$
  &$F_0^{B_s\eta^\prime}(m_{D,D_S}^2)$
  & 0.35 \\
\end{tabular}
\end{ruledtabular}
\end{table}

The $D^0\eta_8$ and $D^0\eta_1$ are not physical final states. The
physical $\eta,\,\eta^\prime$ mesons are defined through
\begin{equation}
\left(
\begin{array}{c}
\eta\\
      \eta^\prime
\end{array}
\right)= \left(
\begin{array}{cc}
\cos\vartheta &-\sin\vartheta\\
\sin\vartheta &\cos\vartheta
\end{array}
\right) \left(
\begin{array}{c}

\eta_8\\
      \eta_1
\end{array}
\right),
\end{equation}
with the mixing angle $\vartheta=-15.4^\circ$
\cite{Feldmann:1998vh}. Form factors are taken from~\cite{LF,MS},
where we list the relevant values in Table~\ref{tab:formfactor}.
For $B_{(s)}\to\eta^{\prime}$ form factors the mixing angle and
Clebsch-Gordan coefficients are included,
\begin{eqnarray}
 F^{B\eta}(m_{D,D_s}^2)&=&\left(
 \frac{\cos\vartheta}{\sqrt6}-\frac{\sin\vartheta}{\sqrt3}\right)F_0^{B\pi}(m_{D,D_s}^2),
 \nonumber\\
 F^{B\eta'}(m_{D,D_s}^2)&=&\left(
 \frac{\sin\vartheta}{\sqrt6}+\frac{\cos\vartheta}{\sqrt3}\right)F_0^{B\pi}(m_{D,D_s}^2)
 \nonumber\\
F^{B_s\eta}(m_{D,D_s}^2)&=&\left(
 -2\frac{\cos\vartheta}{\sqrt6}-\frac{\sin\vartheta}{\sqrt3}\right)F_0^{B_s\eta_s}(m_{D,D_s}^2),
 \nonumber\\
 F^{B_s\eta'}(m_{D,D_s}^2)&=&\left(
 -2\frac{\sin\vartheta}{\sqrt6}+\frac{\cos\vartheta}{\sqrt3}\right)F_0^{B_s\eta_s}(m_{D,D_s}^2),
 \label{eq:FBtoeta}
\end{eqnarray}
where $\eta_s$ is the $s\bar s$ component of $\eta$ and $\eta'$
and the form factor $F^{B\pi}(m_{D,D_s}^2)$ and
$F_0^{B_s\eta_s}(m_{D,D_s}^2)$ are taken from~\cite{LF}
and~\cite{MS}, respectively.

\end{document}